\documentclass[prl,floats,nofootinbib,amssymb,twocolumn,superscriptaddress]{revtex4}
\usepackage{epsf,epsfig}

\usepackage{graphicx}
\usepackage{amssymb}

\newcommand{\be}{\begin{equation}}
\newcommand{\ee}{\end{equation}}
\newcommand{\bea}{\begin{eqnarray}}
\newcommand{\eea}{\end{eqnarray}}
\newcommand{\nn}{\nonumber}

\newcommand{\half}{\frac{1}{2}}

\newcommand{\ba}{\begin{array}}
\newcommand{\ea}{\end{array}}
\newcommand{\bi}{\begin{itemize}}
\newcommand{\ei}{\end{itemize}}
\newcommand{\ben}{\begin{enumerate}}
\newcommand{\een}{\end{enumerate}}

\newcommand{\mb}[1]{\mbox{\boldmath $#1$}}

\preprint{
\hbox to \hsize{
\hfill$\vcenter{\hbox{\bf MADPH-07-14xx}
                \hbox{September 2007}}$}
}

\begin{document}
\title{\vspace*{.5in}
Monochromatic Neutrino Signals from Dark Matter Annihilation}

\author{Vernon~Barger}
\affiliation{Department of Physics, University of Wisconsin, 1150 University
Avenue, Madison, Wisconsin 53706 USA}
\author{Wai-Yee~Keung}
\affiliation{Physics Department, University of Illinois at Chicago, 
Illinois 60607--7059 USA}
\author{Gabe~Shaughnessy}
\affiliation{Department of Physics, University of Wisconsin, 1150 University
Avenue, Madison, Wisconsin 53706 USA}

 

\begin{abstract}
Dark matter (DM) annihilations in the Sun to neutrino-antineutrino pairs have striking signatures in neutrino detectors such as IceCube and KM3.  We make a model independent study of the signals after propagation of the neutrinos from the center of the Sun to the Earth.  A large spin-dependent DM capture cross section in the Sun make the discovery prospects robust.  

\end{abstract}
\thispagestyle{empty}

\maketitle

\section{Introduction}

Dark matter is perhaps the best evidence for physics beyond the standard model (SM).    The amount of dark matter is known to be about 22\% of the total matter-energy budget of the universe, while visible matter only comprises of about 4\% of the universe~\cite{Spergel:2006hy}.  Any realistic model of physics beyond the SM must supply a DM candidate that can reproduce the observed relic density and many models have been suggested that can yield a DM candidate.  Generally, a parity  in the model leads to a stable DM particle.  The leading DM candidate is a neutral weakly interacting stable particle that was produced in the Big Bang.  

Vigorous experimental efforts are underway to identify the nature of DM.  A direct way to identify the DM particle is to observe the recoil of a nuclear target.  A spin-independent (SI) interaction is probed by detecting the recoil of DM from high mass nucleus; the coherent interaction increases the total cross section substantially compared with light nuclei.  The best current limit comes from the XENON10 experiment which found an upper bound of $4.5\times10^{-8}$ pb for a DM mass of $\approx 30$ GeV~\cite{Angle:2007uj}; it is expected that CDMS will soon improve on their current limit of $1.6\times10^{-7}$ pb~\cite{Akerib:2005za}.  The spin-dependent (SD) interaction is governed by the spin of the nucleon.  The upper limits on the SD cross section are 6 orders of magnitude weaker than SI with the best bound from ZEPLIN-II at 0.07 pb~\cite{al.:2007xs}.  For a recent discussions of various direct and indirect DM searches, see Ref.~\cite{DMSAG,Birkedal:2004xn,Cirelli:2005gh}.

At the Large Hadron Collider (LHC), a characteristic signature of the DM particle will be missing transverse energy.  Typically, the DM particle is created at the end of a cascade decay chain of new physics particles.  The reconstruction of the mass of the DM particle at colliders is challenging but feasible.

Additionally, DM is being sought in astrophysics experiments through annihilations to $\gamma$-rays~\cite{Bouquet:1989sr,Stecker:1987dz,Berezinsky:1994wva,Morselli:2002nw,Peirani:2004wy}, antideuterons~\cite{Baer:2005tw} and positrons~\cite{Silk:1984zy,Stecker:1985jc,Tylka:1989xj}  in the galactic halo.

The method of DM detection on which we concentrate in this study is the detection of muon neutrinos from DM annihilations in the Sun and Earth with km$^2$ size neutrino detectors~\cite{Barger:2001ur,Barger:2007xf}.  Currently, the IceCube experiment at the South Pole is underway; it has a muon energy threshold of about 50 GeV~\cite{icecube:2001aa,Halzen:2003fi}.  The planned km$^2$ size detector in the Mediterranean Sea, known as KM3~\cite{KM3:2007xx}, should detect neutrino interactions above a  10 GeV muon energy threshold.

DM is expected to be gravitationally captured in the core of the Sun and Earth and subsequently annihilate.  The rate of capture is strongly dependent on the SD and SI scattering rates that are being limited by direct detection experiments.  For DM capture in the Sun, the SD cross section is the dominant interaction since the Sun is primarily composed of Hydrogen which has a net spin.  The situation for the Earth is very different as the contributions to the capture rate via the SD interaction are negligible for heavy nuclei.  Overall, since the SD cross section is nearly 6 orders of magnitude less constrained than the SI cross section, models that predict large SD rates have greater potential to yield an observable neutrino flux from DM annihilations in the Sun.  

The study of the properties of DM annihilations to neutrino pairs has advantages over a more general neutrino spectrum.  The injected neutrino spectra from this simple process consists of a line at $E_\nu = M_{DM}$~\footnote{Using this injection spectrum, which can be taken as a delta function, one can reproduce any general spectrum after propagation via the Green's function of the evolution equation.  }.  Models in which a neutrino line spectrum from DM annihilations can be realized include a Dirac fermion, Kaluza-Klein spin-1 boson and a scalar that couples to a new vector field.

The solar capture rate of dark matter in the galactic halo is approximately given by \cite{Gould:1992xx}
\bea
C_{\odot}&=& 3.4\times 10^{20} {\rm s}^{-1} \left({\rho_{local}\over 0.3 \text{ GeV/cm}^3}\right) \left({270\text{ km/s}\over v_{local}}\right)^3\nn\\
&&\left({\sigma_{cap}\over 10^{-6}\text{ pb}}\right)\left({100\text{ GeV}\over M_{DM}}\right)^2,
\label{eq:scaprate}
\eea
where $\rho_{local}$ and $v_{local}$ are the local density and velocity of relic dark matter, respectively.  The average density of $\rho_{local} \approx 0.3  \text{ GeV/cm}^3$ may be enhanced due to caustics in the galactic plane~\cite{Sikivie:2001fg}.  The effective strength of the capture cross section of DM with solar matter is given by $\sigma_{cap} = \sigma_{SD}^H + \sigma_{SI}^{H} + 0.07 \sigma_{SI}^{He}$.  The factor of $0.07$ before $\sigma_{SI}^{He}$ comes from the relative abundance of helium and hydrogen in the Sun, as well as other dynamical and form factor effects~\cite{Halzen:2005ar}.  The cross sections determine how efficiently the Sun slows and captures DM.  The value of $\sigma_{cap}$ has considerable variation with models.  


The capture rate of dark matter in the galactic halo by the Earth  is approximately given by \cite{Jungman:1995df}
\bea
C_{\oplus}&=& 4.8\times10^{13} {\rm s}^{-1} \left({\rho_{local}\over 0.3 \text{ GeV/cm}^3} \right) f_\oplus(M_{DM})\nn\\
&& \left({\sigma_{cap}\over 10^{-6}\text{ pb}}\right) \left({  \mu_{p-DM} \over1 \text{ GeV} }\right)^{-2}\label{eq:ecaprate},
\eea
where $\mu_{p-DM}={M_{DM} m_p\over M_{DM}+m_p}$ is the reduced mass of the DM-nucleon system and $\sigma_{cap} = \sigma_{SI}^H$.  The form factor, $f_\oplus(M_{DM})$, is given in Ref. \cite{Jungman:1995df}.  A positive signal from the Earth, but not the Sun would strongly suggest that the DM particle has no significant SD interaction with matter.

\section{Neutrino production and propagation}
\label{sect:nuprop}

We assume that DM annihilates at the capture rate to neutrino pairs with a democratic ratio of $\nu_e:\nu_\mu:\nu_\tau:\bar\nu_e:\bar\nu_\mu:\bar\nu_\tau = 1:1:1:1:1:1$, producing a characteristic line shape in the initial neutrino energy spectra.  The neutrinos are then propagated through the Sun as in Ref.~\cite{Barger:2007xf}.  The evolution equation is given by~\cite{Cirelli:2005gh}
\be
\label{eq:prop}
D_{\rm prop} {\mb \rho} \equiv {d {\mb\rho} \over dr}  +i [{\mb H},{\mb\rho}] - \left.{d {\mb\rho}\over dr}\right|_{NC} - \left.{d {\mb\rho}\over dr}\right|_{CC} - \left.{d {\mb\rho}\over dr}\right|_{inj}=0
\ee
where $\mb\rho(r,E)$ is the complex density matrix in the gauge eigenbasis describing the state of the neutrino of energy $E$ at a distance $r$ from the center of the Sun.   The Hamiltonian, $\mb H$, includes the effects of vacuum oscillation from nonzero neutrino mass splitting and the matter interaction:
\be
{\mb H} = {{\mb m}^\dagger {\mb m}\over 2E_\nu}+\sqrt{2} G_F \left[ N_e(r) \delta_{i1}\delta_{j1}-{N_n(r)\over2}\delta_{ij}\right].
\label{eq:ham}
\ee
Here $\mb m$ is the neutrino mass matrix in the gauge eigenstate basis, $E_\nu$ is the neutrino energy, $G_F=1.66\times10^{-5}\text{ GeV}^{-2}$ is the Fermi constant and $N_e(r)$ and $N_n(r)$ are the electron and neutron densities in the Sun~\cite{Bahcall:2004yr}.  The Neutral Current (NC) and Charged Current (CC) source terms describe the absorption and re-injection of neutrinos caused by NC and CC processes while the injection term describes the spectra injected from DM annihilation in the core of the Sun or Earth

\section{Muon rate at IceCube}
\label{sect:icecube}

We calculate the detection rates with up-going muons following the simulation for IceCube outlined in Ref. \cite{GonzalezGarcia:2005xw}.  The neutrino flux at the surface of the Earth is given by
\be
\label{eq:flux}
{d\Phi_\nu \over dE_\nu} = {1 \over 2} {C \over 4 \pi R^2}{1\over N}{d N \over d E_\nu} \text{BF}(DM DM \to \nu \bar \nu) ,
\ee
where the parameter $C$ is the DM capture rate in the Sun ($C_\odot$) or Earth ($C_\oplus$) given explicitly above in Eqs. (\ref{eq:scaprate}) and (\ref{eq:ecaprate}), respectively.  The factor of $\half$ is associated with the fact that two DM particles produce one annihilation event; here $R=1.49\times10^{11}{\rm m}$ is the Earth-Sun distance for annihilation in the Sun, or $R=6.4\times10^6$ m is the radius of the Earth for annihilation in the Earth's core.  The first factor in Eq. (\ref{eq:flux}) describes the flux at the surface of the Earth while the ${1\over N}{d N \over d E_\nu} $ term is the normalized differential rate and $\text{BF}(DM DM \to \nu \bar \nu)$ is the fraction of annihilations to neutrino-antineutrino pairs.

The muon rate in a km$^2$ area detector such as IceCube can be determined by folding the neutrino flux with the muon production cross section~\cite{Halzen:2005ar,Barger:2007xf}
\bea
{d N_\mu\over dE_\mu} &=& \int_{E_\mu}^{\infty} {d\Phi_{\nu_\mu} \over dE_{\nu_{\mu}}} \left[{d \sigma^p_\nu (E_{\nu_\mu}, E_\mu)\over dE_\mu} \rho_{p}+{d \sigma^n_\nu (E_{\nu_\mu}, E_\mu)\over dE_\mu} \rho_{n}\right] \nn\\
&&R_\mu(E_\mu) A_{eff}\left(E_\mu\right) dE_{\nu_\mu} + \left(\nu_\mu \to \bar \nu_\mu\right).
\label{eq:numurate}
\eea
The densities of protons and neutrons near the detector are $\rho_{p} = {5\over 9}N_A \text{ cm}^{-3}$ and $\rho_{n} = {4\over 9}N_A \text{ cm}^{-3}$, respectively, where $N_A$ is Avagadro's number\footnote{Since the muon range is at most 1 km for a 1 TeV muon, the point of muon production can be assumed to be in ice, rather than the Earth's crust.}.  The muon range, $R_\mu(E_\mu)$, is the distance a muon travels before its energy falls below a threshold energy, $E_\mu^{thr}$.   We take $E_\mu^{thr} = 50$ GeV, which is optimistic for IceCube, but the muon thresholds are expected greatly improve~\cite{fhalzen}.  Due to the long muon range, the fiducial volume of the detector can be factored as the product of the range and the cross sectional area of the detector, called the effective area.  The effective area of the detector, $A_{eff}$, is calculated for IceCube following Ref.~\cite{GonzalezGarcia:2005xw}.

We apply Eq. (\ref{eq:numurate}) to calculate neutrino signals in IceCube from DM annihilations and the backgrounds from atmospheric neutrinos.  The backgrounds from atmospheric neutrinos are calculated following Ref.~\cite{Barger:2007xf}.

\begin{figure}
   \begin{center}\vspace{.1in}
      \includegraphics[width=0.45\textwidth]{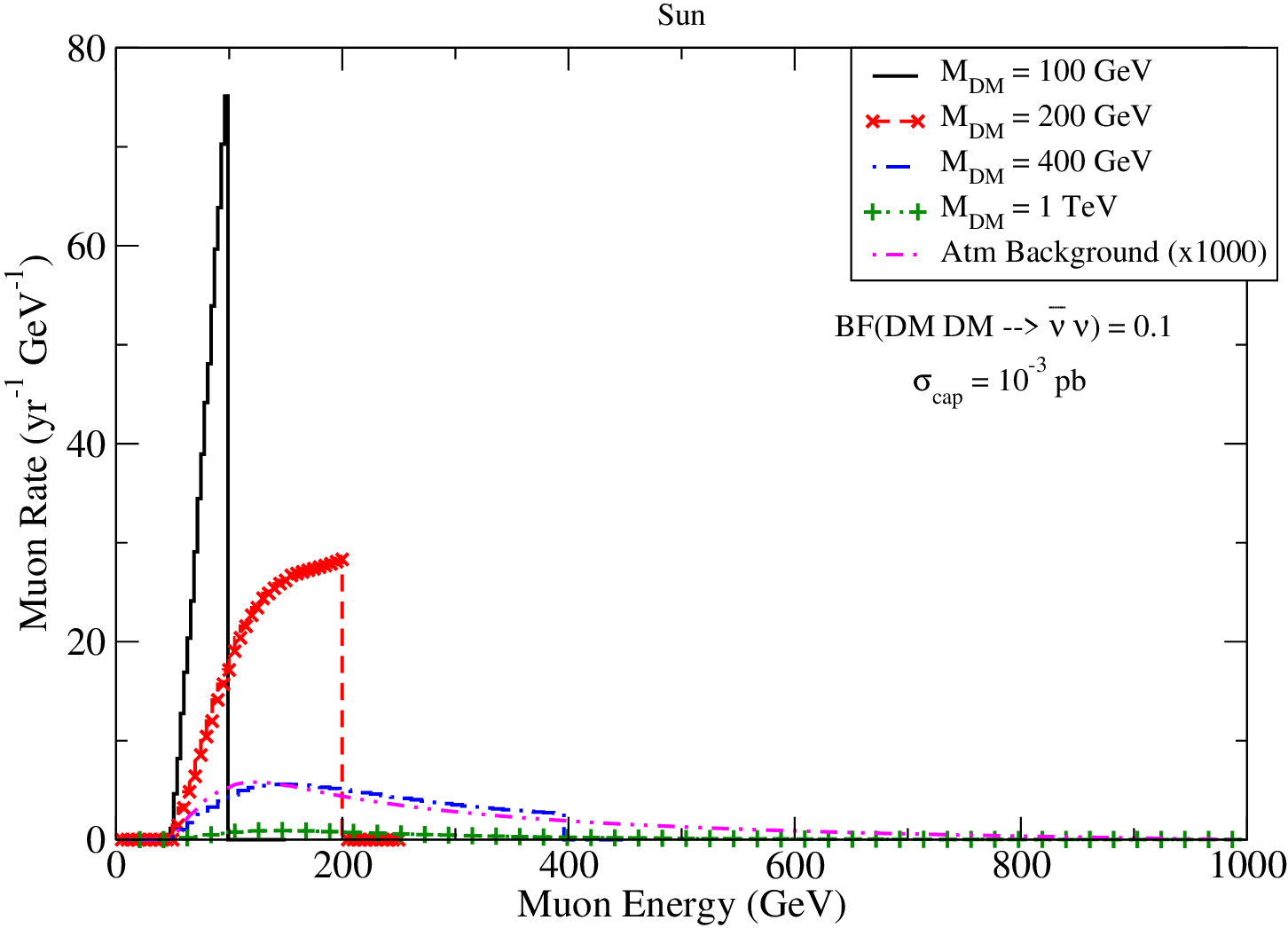}\vspace{.30in}
      \includegraphics[width=0.45\textwidth]{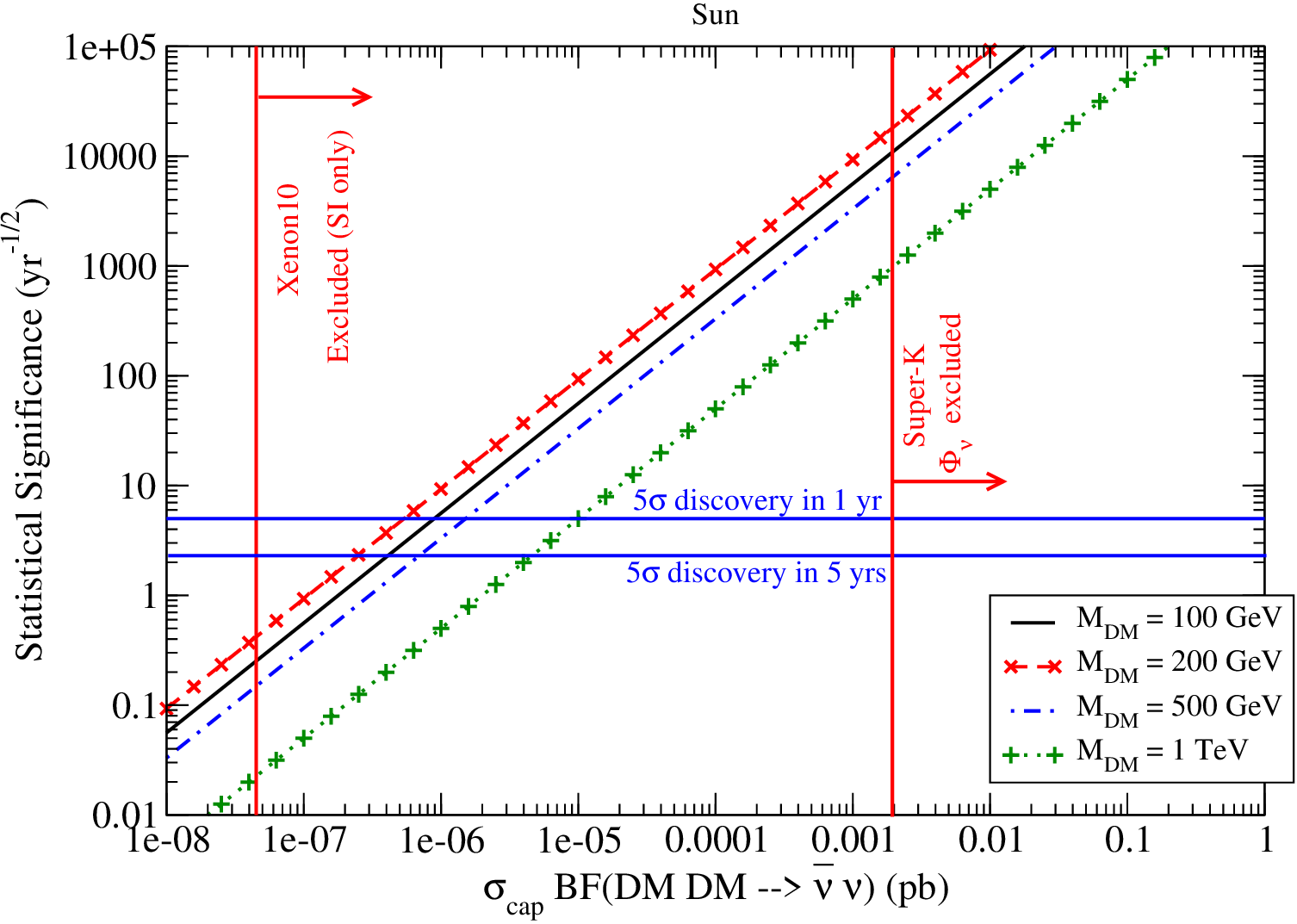}
      \caption{(a) Predicted muon event rates in IceCube from monochromatic neutrinos.  The muon rates assume $\text{BF}(DM DM \to \nu \bar \nu) = 0.1$ and a capture cross-section $\sigma_{cap} = 10^{-3}$ pb that is dominated by the SD interaction.  The signal region includes $E_\mu < 300$ GeV.  (b) Statistical significance for discovery.  The discovery region is given above the blue horizontal lines at $5\sigma$ for one and five years of data.  The limits from Xenon10 conservatively assume the  maximum value of $\text{BF}(DM DM \to \nu \bar \nu)=1$ while the Super-Kamiokande exclusion is a direct measurement limit on the high energy muon-neutrino flux.}
      \label{fig:sun}
   \end{center}
\end{figure}

{\it Annihilations in the Sun}:  The DM accumulation and annihilation rates in the Sun are assumed to equilibrate: $\Gamma_{ann}=\half C_{\odot}$.  In Fig. \ref{fig:sun}a, we show the signal muon event rate and atmospheric background~\cite{Honda:2006qj} in IceCube for four DM masses.  The numbers of predicted signal events, based on a SD capture cross section of $\sigma_{cap} = 10^{-3}$ pb and a branching fraction of 10\%, for $M_{DM}$ values of 100, 200, 400 and 1000 GeV are 1800, 3000, 11000 and 160 events per year, respectively.  The atmospheric background for the 50 - 300 GeV window is 10.4 events per year.

We show the statistical significances in Fig. \ref{fig:sun}b for a broad energy region of $E_{\mu}^{thr} < E_\mu < 300$ GeV.  This choice of the upper cut at 300 GeV is because the large attenuation of high energy neutrinos by the CC and NC interactions in the Sun suppresses signal contributions at higher energies.  Once evidence is seen for a signal, the window can be narrowed to improve the significance.

Given the restrictive experimental limit on $\sigma_{SI}$, there is no hope to see a corresponding signal from annihilation to neutrino pairs in the Sun.  The SD cross section must be large for IceCube to probe the $\nu$-line signal.  The Super-Kamiokande experimental limit on the induced muon flux is approximately $5\times 10^{-15}$ cm$^{-2}$ s$^{-1}$~\cite{Desai:2004pq}, and is given by the vertical line in Fig. \ref{fig:sun}b.

Information on the nature of the DM can be extracted from the muon spectra of the signal shown in Fig. \ref{fig:sun}.  A determination of the DM mass can be made from the muon energy distribution and the capture rate of DM in the Sun can be determined from the total signal rate.  The capture rate in turn gives information on the local galatic DM density and its velocity (cf. Eq. (\ref{eq:scaprate})) that can complement inferences from direct detection experiments.  Additional properties of DM can be extracted by measuring the neutrino flavor ratios~\cite{Lehnert:2007fv}.

Shown in Fig.~\ref{fig:sun}b is the statistical significance (with one year of data) of a signal relative to the atmospheric neutrino background vs $ \sigma_{cap} \times \text{BF}(DM DM \to \nu \bar \nu)$ for various DM masses.  The $5\sigma$ discovery regions for IceCube are above the blue horizontal lines.  The limit from Xenon10 of $4.5\times 10^{-8}$ pb conservatively assumes $\text{BF}(DM DM \to \nu \bar \nu)=1$ and that the capture rate is entirely given by the SI interaction (as in the case of scalar DM with no new gauge fields).  Therefore, a model which has no SD interaction is excluded to the right of the Xenon10 line.  For values of $\text{BF}(DM DM \to \nu \bar \nu)<1$, the vertical exclusions lines move to the left.  It is evident that the prospects for IceCube to discover  DM annihilations to neutrinos from the solar core are robust for the SD case, but are dim for the SI case.



{\it Annihilations in the Earth:}  We optimistically assume that $\Gamma_{ann} = \half C_{\oplus}$ for which the equilibration time is longer than the age of the solar system. Otherwise, $\Gamma = \half C_{\oplus} \tanh^2\left(\sqrt{C_{\oplus} A_{\oplus}}t_{Earth}\right)$, where $A_{\oplus}={\langle \sigma v\rangle\over V_{\oplus}}$ is the annihilation rate times relative velocity per unit volume and $t_{Earth}=4.5$ Gy.  However, equilibrium may not have yet been reached in the case of the Earth~\footnote{The equilibration time for the Earth is expected to be approximately a thousand times longer than that of the Sun~\cite{Jungman:1995df}.}.  If not in equilibrium, the annihilation rate is dependent on the DM annihilation cross section.  By assuming the DM capture and annihilation processes are in equilibrium, we are assuming the best-case scenario.

The angular area of acceptance is larger than the solar case due to the less concentrated density profile of the dark matter in the Earth.  It is estimated~\cite{Jungman:1995df} that about 98\% of the DM is contained within a window of 
\be
\Delta \theta = 6.3^\circ \left({100\text{ GeV}\over M_{DM} }\right)^{1/2}
\ee
about $\theta_z = 180^\circ$.  Thus, the statistical significance of the annihilation signal improves with larger $M_{DM}$.  

\begin{figure}[t]
   \begin{center}
      \includegraphics[width=0.45\textwidth]{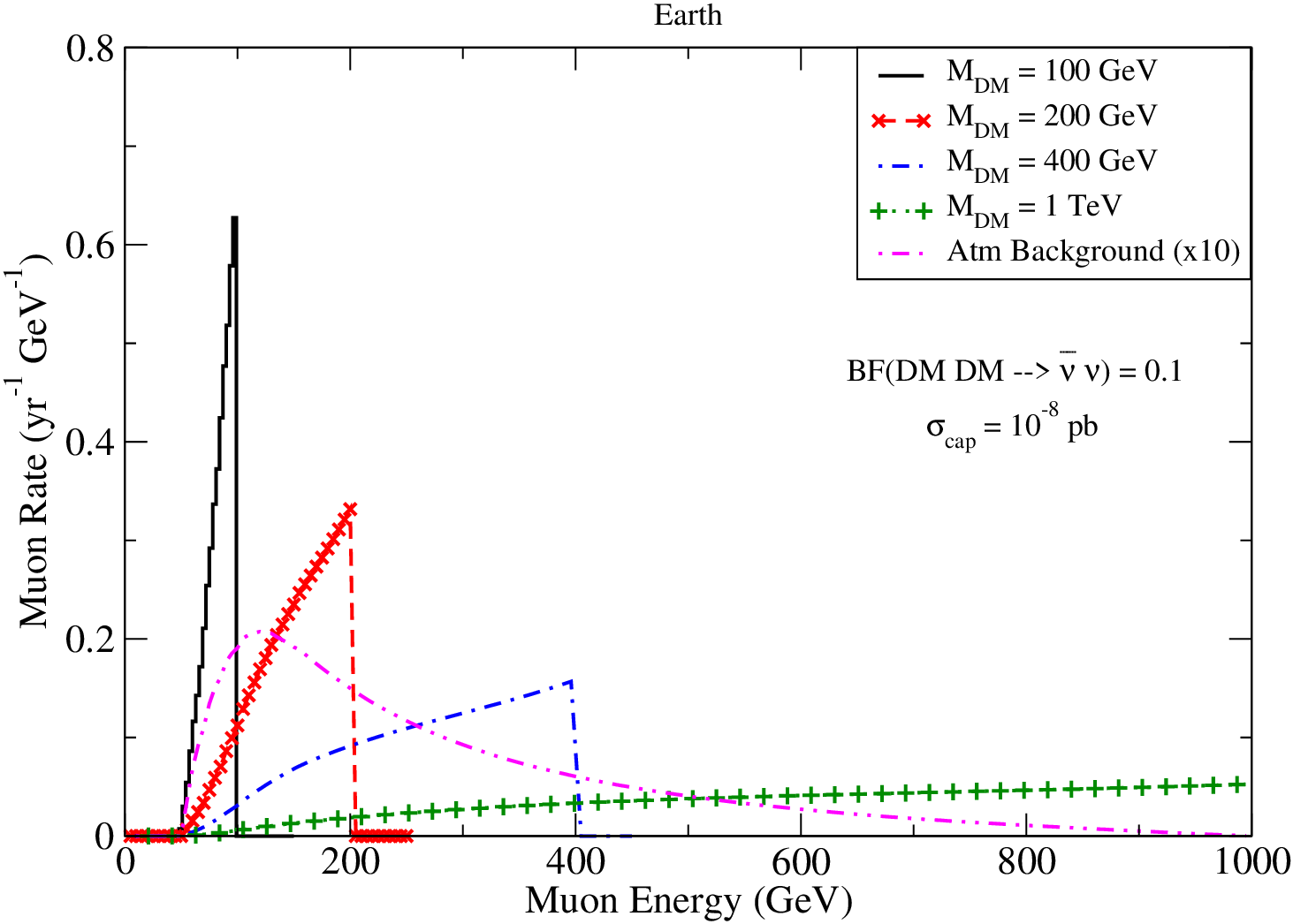}\vspace{.4in}
      \includegraphics[width=0.45\textwidth]{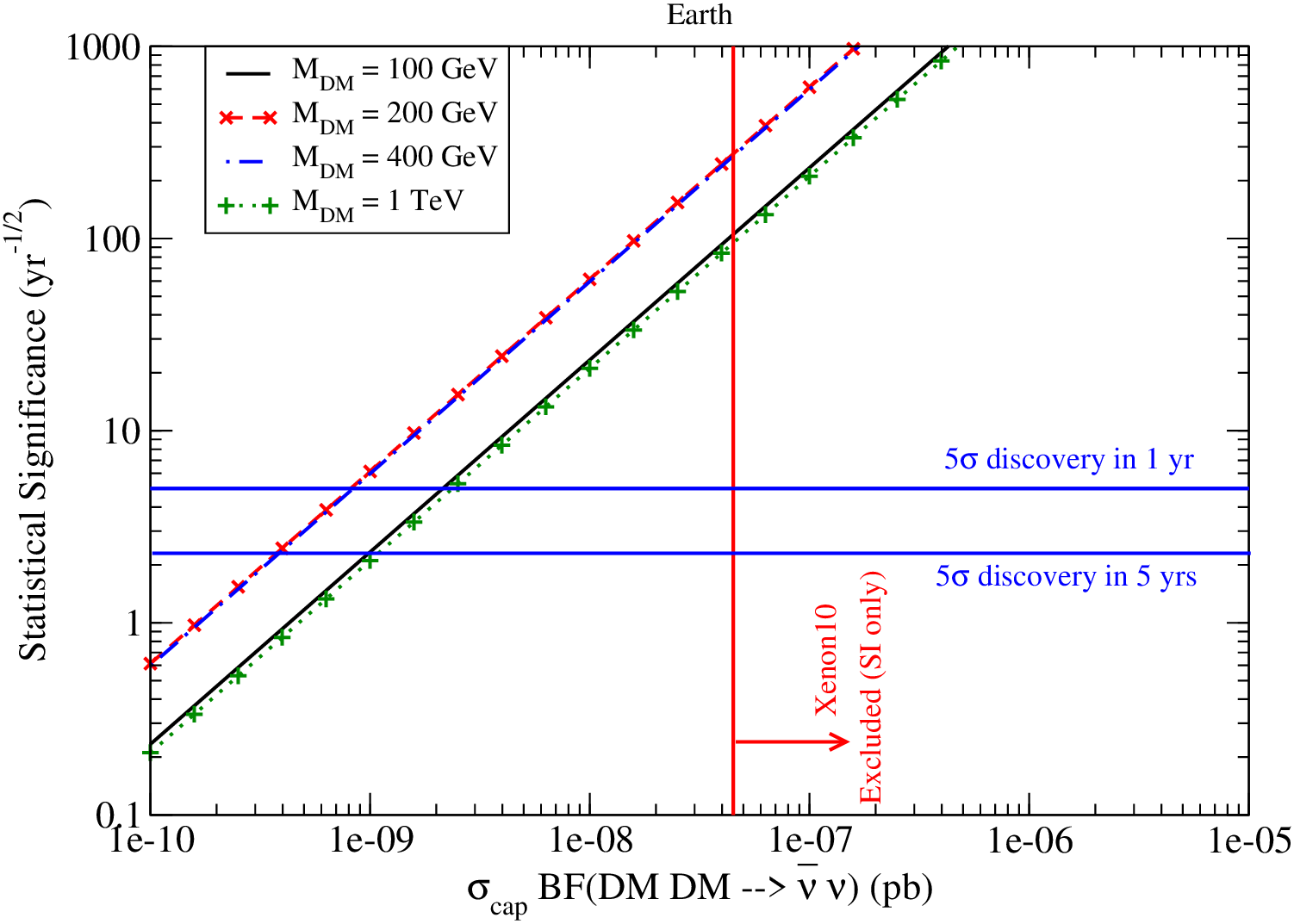}
      \caption{Similar to Fig. \ref{fig:sun}, but for the Earth where the SI interaction dominates the capture rate.  The muon rate is illustrated with a SI capture cross-section of $\sigma_{cap} = 10^{-8}$ pb and a $DM DM \to \bar \nu \nu$ branching fraction of 10\%.}
      \label{fig:earth}
   \end{center}
\end{figure}

In Fig. \ref{fig:earth}a, we show the expected muon rates in IceCube for $\sigma_{cap} = 10^{-8}$ pb and $\text{BF}(DM DM\to \bar \nu \nu) = 0.1$.  Constraints from Super Kamiokande, with $\Phi_\nu \lesssim 3\times10^{-14}\text{ cm}^{-2}\text{s}^{-1}$~\cite{Desai:2004pq}, are not strong enough to limit the possibility from a signal from the Earth which we predicted to be about $\Phi_\nu \approx 10^{-19}\text{ cm}^{-2}\text{s}^{-1}$.  The signal is substantially greater than the background; the numbers of predicted signal events based on a SI capture cross section for $M_{DM}$ values of 100, 200, 400 and 1000 GeV are 1.4, 2.6, 1.8 and 0.4 events per year, respectively.   The signal is enhanced at $M_{DM}\sim 60 $ GeV due to the abundance of Iron in the Earth~\cite{Jungman:1995df}.  The high IceCube muon energy threshold of 50 GeV limits the opportunity of discovery.  For the KM3 detector~\cite{KM3:2007xx}, with a lower energy threshold, the prospects for discovery may be more likely.

In summary, the search for DM annihilations in the Sun to neutrino pairs is very promising, provided that the SD interaction dominates the DM capture rate.  The case for corresponding discovery of DM annihilations in the Earth is less promising due to the strong limits on the capture rate because of the stringent upper bound on the SI cross-section.

\begin{acknowledgments}
\end{acknowledgments}
We thank S. Desai, F. Halzen and T. Weiler for valuable discussions.  This work was supported in part by the U.S.~Department of Energy under grant Nos. DE-FG02-95ER40896 and DE-FG02-84ER40173, and by the Wisconsin Alumni Research Foundation.

\bibliographystyle{h-physrev}
\bibliography{nuline}                      

\begin{thebibliography}{10}

\bibitem{Spergel:2006hy}
WMAP, D.~N. Spergel {\em et~al.},
\newblock Astrophys. J. Suppl. {\bf 170}, 377 (2007).

\bibitem{Angle:2007uj}
J.~Angle {\em et~al.},
\newblock (2007), arxiv:0706.0039.

\bibitem{Akerib:2005za}
D.~S. Akerib {\em et~al.},
\newblock Phys. Rev. {\bf D73}, 011102 (2006).

\bibitem{al.:2007xs}
G.~J. A.~e. al.,
\newblock (2007), arXiv:0708.1883.

\bibitem{DMSAG}
Dark Matter Scientific Assessment Group , http://www.science.doe.gov/hep.

\bibitem{Birkedal:2004xn}
A.~Birkedal, K.~Matchev, and M.~Perelstein,
\newblock Phys. Rev. {\bf D70}, 077701 (2004).

\bibitem{Cirelli:2005gh}
M.~Cirelli {\em et~al.},
\newblock Nucl. Phys. {\bf B727}, 99 (2005), hep-ph/0506298.

\bibitem{Bouquet:1989sr}
A.~Bouquet, P.~Salati, and J.~Silk,
\newblock Phys. Rev. {\bf D40}, 3168 (1989).

\bibitem{Stecker:1987dz}
F.~W. Stecker,
\newblock Phys. Lett. {\bf B201}, 529 (1988).

\bibitem{Berezinsky:1994wva}
V.~Berezinsky, A.~Bottino, and G.~Mignola,
\newblock Phys. Lett. {\bf B325}, 136 (1994).

\bibitem{Morselli:2002nw}
GLAST, A.~Morselli, A.~Lionetto, A.~Cesarini, F.~Fucito, and P.~Ullio,
\newblock Nucl. Phys. Proc. Suppl. {\bf 113}, 213 (2002).

\bibitem{Peirani:2004wy}
S.~Peirani, R.~Mohayaee, and J.~A. de~Freitas~Pacheco,
\newblock Phys. Rev. {\bf D70}, 043503 (2004).

\bibitem{Baer:2005tw}
H.~Baer and S.~Profumo,
\newblock JCAP {\bf 0512}, 008 (2005).

\bibitem{Silk:1984zy}
J.~Silk and M.~Srednicki,
\newblock Phys. Rev. Lett. {\bf 53}, 624 (1984).

\bibitem{Stecker:1985jc}
F.~W. Stecker, S.~Rudaz, and T.~F. Walsh,
\newblock Phys. Rev. Lett. {\bf 55}, 2622 (1985).

\bibitem{Tylka:1989xj}
A.~J. Tylka,
\newblock Phys. Rev. Lett. {\bf 63}, 840 (1989).

\bibitem{Barger:2001ur}
V.~D. Barger, F.~Halzen, D.~Hooper, and C.~Kao,
\newblock Phys. Rev. {\bf D65}, 075022 (2002).

\bibitem{Barger:2007xf}
V.~Barger, W.-Y. Keung, G.~Shaughnessy, and A.~Tregre,
\newblock (2007), arXiv:0708.1325 [hep-ph].

\bibitem{icecube:2001aa}
J.~Ahrens {\em et~al.},
\newblock (2001), IceCube Preliminary Design Document.

\bibitem{Halzen:2003fi}
F.~Halzen and D.~Hooper,
\newblock JCAP {\bf 0401}, 002 (2004).

\bibitem{KM3:2007xx}
http://www.km3net.org/.

\bibitem{Gould:1992xx}
A.~Gould,
\newblock Astrophys. J. {\bf 388} (1992).

\bibitem{Sikivie:2001fg}
P.~Sikivie,
\newblock Phys. Lett. {\bf B567}, 1 (2003).

\bibitem{Halzen:2005ar}
F.~Halzen and D.~Hooper,
\newblock Phys. Rev. {\bf D73}, 123507 (2006).

\bibitem{Jungman:1995df}
G.~Jungman, M.~Kamionkowski, and K.~Griest,
\newblock Phys. Rept. {\bf 267}, 195 (1996).

\bibitem{Bahcall:2004yr}
J.~N. Bahcall, S.~Basu, M.~Pinsonneault, and A.~M. Serenelli,
\newblock Astrophys. J. {\bf 618}, 1049 (2005).

\bibitem{fhalzen}
F.~Halzen, private communication.

\bibitem{GonzalezGarcia:2005xw}
M.~C. Gonzalez-Garcia, F.~Halzen, and M.~Maltoni,
\newblock Phys. Rev. {\bf D71}, 093010 (2005).

\bibitem{Honda:2006qj}
M.~Honda, T.~Kajita, K.~Kasahara, S.~Midorikawa, and T.~Sanuki,
\newblock Phys. Rev. {\bf D75}, 043006 (2007).

\bibitem{Desai:2004pq}
S.~Desai {\em et~al.},
\newblock Phys. Rev. {\bf D70}, 083523 (2004), hep-ex/0404025.

\bibitem{Lehnert:2007fv}
R.~Lehnert and T.~J. Weiler,
\newblock (2007), arXiv:0708.1035 [hep-ph].

\end{thebibliography}


\end{document}